\documentclass[preprint,12pt]{elsarticle}

\usepackage{graphics}
\usepackage{graphicx}
\usepackage{epstopdf}
\usepackage{amssymb}
\journal{Journal of Molecular Structure}
\def\re#1{(\ref{#1})}
\def\be{\begin{equation}}
\def\ee{\end{equation}}
\def\bea{\begin{eqnarray}}
\def\eea{\end{eqnarray}}
\def\bra#1{ \langle #1 |}
\def\ket#1{|#1\rangle}

\def\cm{cm$^{-1}\,$}
\def\ms{\nu_{\rm OH}}
\def\mb{\delta_{\rm OH}}
\def\go{\gamma_{\rm 1}}
\def\gt{\gamma_{\rm 2}}
\def\hb{\rm HB}

\begin{document}
\begin{frontmatter}
\title{IR Spectrum of the O-H$\cdots$O Hydrogen Bond of Phthalic Acid Monomethylester in Gas Phase and in CCl$_4$ Solution}
\author[UR]{Yun-an Yan}
\author[FU,BG]{M. Petkovi\'{c}}
\author[FU]{Gireesh M. Krishnan}
\author[UR]{Oliver K\"uhn\corref{cor1}}
\address[UR]{Institut f\"ur Physik, Universit\"at Rostock, D-18051 Rostock, Germany}
\address[FU]{Institut f\"ur Chemie und Biochemie, Freie Universit\"at Berlin, Takustr. 3, D-14195 Berlin, Germany}
\address[BG]{Faculty of Physical Chemistry, University of Belgrade, Studentski trg 12-16, 11158 Belgrade, Serbia}
\cortext[cor1]{Corresponding author}
\ead{oliver.kuehn@uni-rostock.de}
\begin{abstract}
The absorption spectrum of  the title compound  in the spectral range of the Hydrogen-bonded OH-stretching vibration has been investigated using a five-dimensional gas phase  model as well as a QM/MM classical molecular dynamics simulation in solution. The gas phase model predicts a Fermi-resonance  between the OH-stretching fundamental and the first OH-bending overtone transition with considerable oscillator strength redistribution. The anharmonic coupling to a low-frequency vibration of the Hydrogen bond leading to a vibrational progression is studied within a diabatic potential energy curve model. The condensed phase simulation of the dipole-dipole correlation function results in a broad band in the 3000 \cm region in good agreement with experimental data. Further, weaker absorption features around 2600 \cm have been identified as being due to motion of the Hydrogen within the Hydrogen bond.
\end{abstract}


\end{frontmatter}
\newpage
%
\section{Introduction}
\label{sec:intro}
%
Infrared (IR) spectroscopic studies of the  dynamics of Hydrogen bonds (HBs) continue to trigger considerable theoretical and experimental efforts not at least due to the outstanding importance of this type of bonding for a variety of phenomena in chemical and biological physics \cite{hynes06,marechal07}. In recent years ultrafast IR spectroscopy has demonstrated its capability to unravel the details of the dynamics hidden in linear absorption spectra \cite{nibbering04:1887}.
Phthalic acid monomethylester (PMME) has served as model system for studying the ultrafast vibrational dynamics of medium strong HBs \cite{nibbering07:619,giese06:211,stenger02_256,stenger01:2929,madsen02_909,heyne04:6083,kuhn02:7671}. Experimental results include the observation of wave packet dynamics due to excitation of a low-frequency mode  which modulates the HB length and which is coupled to the OH-stretching fundamental transition \cite{stenger01:2929,madsen02_909}, and the unraveling of the path for the ultrafast vibrational energy relaxation using two-color pump-probe spectroscopy \cite{heyne04:6083}. The observed subpicosecond relaxation of the OH-stretching mode has been explained by an energy cascading mechanism involving different OH-bending modes of the HB. 
Concerning the linear IR spectrum, however, there is only a study for the deuterated case which combined a three-dimensional model Hamiltonian with rates for phase and energy relaxation according to dissipation theory \cite{kuhn02:7671}. Specifically, it was found that the IR lineshape in the OD-stretching region is dominated by a Fermi-resonance between the OD-stretching fundamental and the OD-bending overtone. The resulting double peak shape of the absorption is typical for Hydrogen-bonded systems, see Refs. \cite{bratos02:5,rousseau02:241}.

The present study sets the focus on the IR spectrum of the normal species of PMME in the range of the OH-stretching fundamental transition around 3000 \cm. Two complementary models will be considered. First, a five-dimensional gas phase Hamiltonian is determined and diagonalized to obtain detailed insight into the composition of vibrational eigenstates in the considered spectral range. Degrees of freedom are selected starting from the OH-stretching and bending modes which are supplemented by additional modes according to calculated force constants up to fourth order. Second, hybrid quantum mechanics/molecular mechanics (QM/MM) simulations of classical trajectories of PMME in CCl$_4$ solution are performed. Besides an analysis of the HB geometry in solution, the dipole-dipole autocorrelation function is calculated from which the IR absorption spectrum is obtained (for related applications to Hydrogen-bonded systems, see Refs. \cite{gaigeot03:10344,rousseau04:4804,sauer05:1706,vener06:273,jezierska07:818}). 

The paper is organized as follows: Section \ref{sec:methods} starts with an account on the determination of the Hamiltonian for PMME in gas phase. This is followed by a summary of the QM/MM protocol. Results for the gas phase IR spectrum are given in Section \ref{sec:res:gas} and the analysis of the condensed phase simulations is presented in Section \ref{sec:res:cond}. The paper is summarized in Section \ref{sec:sum}.
%
\section{Theoretical Methods}
\label{sec:methods}
%
%
\subsection{Anharmonic Potential Energy Surface in Gas Phase}
\begin{figure}
\centering
\includegraphics[width=0.7\textwidth]{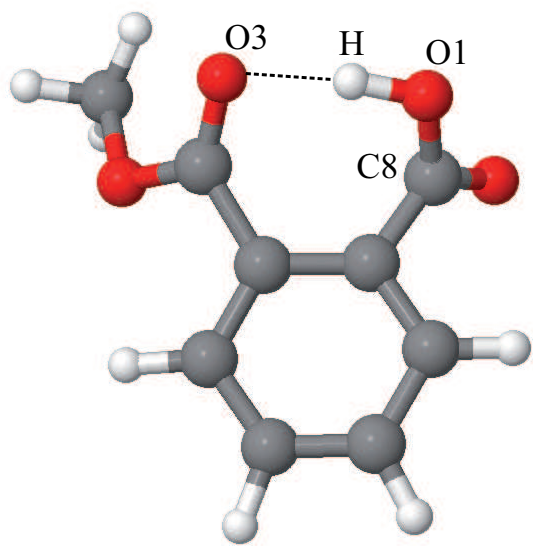}
\caption{Structure of phthalic acid monomethylester (PMME) as obtained by geometry optimization using the DFT/B3LYP level
of theory with a Gaussian 6-31+G(d,p) basis set. In the analysis of the trajectory
 the angles $\alpha=\angle(\textrm{C}8,\textrm{O}1,\textrm{H})$, 
$\beta=\angle(\textrm{O}1,\textrm{H},\textrm{O}3)$, and the dihedral angle $\phi$
formed by the atoms $\textrm{C}8$, $\textrm{O1}$, H, and $\textrm{O}3$ will be used.
}
\label{fig:struct}
\end{figure}

%
%
The electronic ground state geometry of PMME in the gas phase has been optimized using the DFT/B3LYP level
of theory with a Gaussian 6-31+G(d,p) basis set \cite{gaussian98}, see Fig. \ref{fig:struct}. A detailed account on the gas phase geometry as well as on its dependence on the quantum chemical method can be found in Ref.  \cite{paramonov01:205}. Focussing on the  vicinity of the most stable configuration we have chosen to express the gas phase  Hamiltonian in terms of normal mode coordinates $\{Q_i\}$. In general one can write the PES in terms of a correlation expansion \cite{carter97:10458}
\bea
V(\{Q_i\}) &=&\sum_i V_i^{(1)}(Q_i) + \sum_{i<j} V^{(2)}_{ij}(Q_i,Q_j) +  \sum_{i<j<k} V^{(3)}_{ijk}(Q_i,Q_j,Q_k) \nonumber\\
&+&  \sum_{i<j<k<l} V^{(4)}_{ijkl}(Q_i,Q_j,Q_k,Q_l) + \ldots 
\eea
Restricting to four-mode correlations this type of  PES is readily obtained by the anharmonic expansion
\be
V(\{Q_i\}) = \frac{1}{2} \sum_i \omega_i^2 Q_i^2 + \frac{1}{3!} \sum_{ijk} K_{ijk} Q_iQ_jQ_k + \frac{1}{4!} \sum_{ijkl} K_{ijkl} Q_iQ_jQ_kQ_{l} \, .
\label{taylor}
\ee
Here, the $\omega_i$ are the harmonic frequencies and the  $K_{ijk}$ and $K_{ijkl}$ are the third- and fourth-order anharmonic couplings. For the present case of PMME they have been calculated by combining finite differencing with analytical second derivatives as described in Refs. \cite{schneider89:367,csaszar98:13}. 

\begin{figure}
\begin{center}
\includegraphics[width=\textwidth]{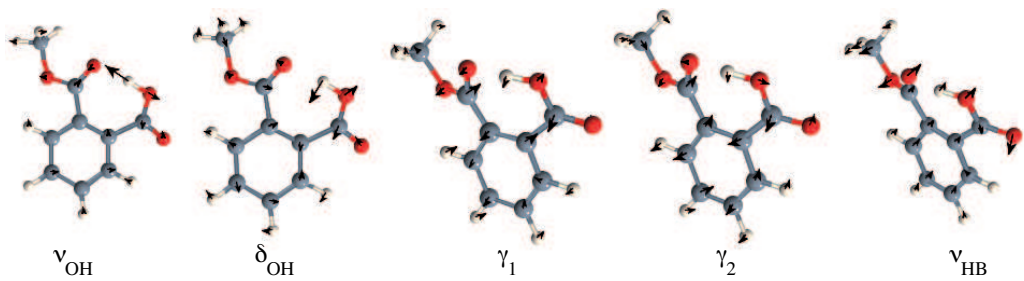}
\caption{Displacement vectors for the gas phase normal modes which define the 5D model.}
\label{fig:modes}
\end{center}
\end{figure}

Being interested in the PMME HB dynamics we have selected a set of relevant modes comprising a reduced five-dimensional (5D) quantum mechanical model as follows: Starting point are the coupled OH-stretching ($\ms$) and  OH-bending ($\mb$)  modes with harmonic frequencies of 3279 \cm and 1446 \cm, respectively, shown in Fig. \ref{fig:modes}.  Further we included the strongest coupled low-frequency mode $\nu_{\hb}$ with harmonic frequency of 39 \cm. 
As can be seen from Fig. \ref{fig:modes} excitation of this mode leads to a periodic modulation of the O-O distance in the HB. Hence the choice of such a mode is motivated by the experimental observation of damped low-frequency wave packet motion  in Ref. \cite{stenger01:2929}. Note, however, that from the oscillatory component of the pump-probe signal a frequency of $\sim$100 \cm had been extracted. Looking for higher frequency modes of similar character one finds a vibration with a harmonic frequency of 72 \cm \cite{stenger01:2929}. However, its anharmonic coupling to the stretching vibration is considerably smaller than for the selected $\nu_{\hb}$. Further, for such low-frequency modes the harmonic approximation performs usually rather poor. Indeed the anharmonic frequency of $\nu_{\hb}$ within the present model is 60 \cm.
The selection of further modes is guided by the available anharmonic coupling constants as well as the possible resonances.  Here, we find two  modes, $\go$ and $\gt$, which have out-of-plane OH-bending character (see, Fig. \ref{fig:modes}) that could have significant influence on the
OH-stretching and in particular OH-bending vibration. Their harmonic frequencies are 682 \cm and 785 \cm, respectively, and  their combination as well as the overtone transitions are close in resonance to the fundamental transition of mode $\mb$. Moreover, the combination of $\mb$ and the overtones of $\go$ and $\gt$ is close in energy to the OH-stretching fundamental transition. The most important anharmonic coupling constants for these five modes are summarized in Tab. \ref{tab:afc}. The 5D model coordinates will be labeled as $\{Q_{\rm s},Q_{\rm b},Q_{\go},Q_{\gt},Q_{\hb}\}$.

\begin{figure}
\begin{center}
\includegraphics[width=\textwidth]{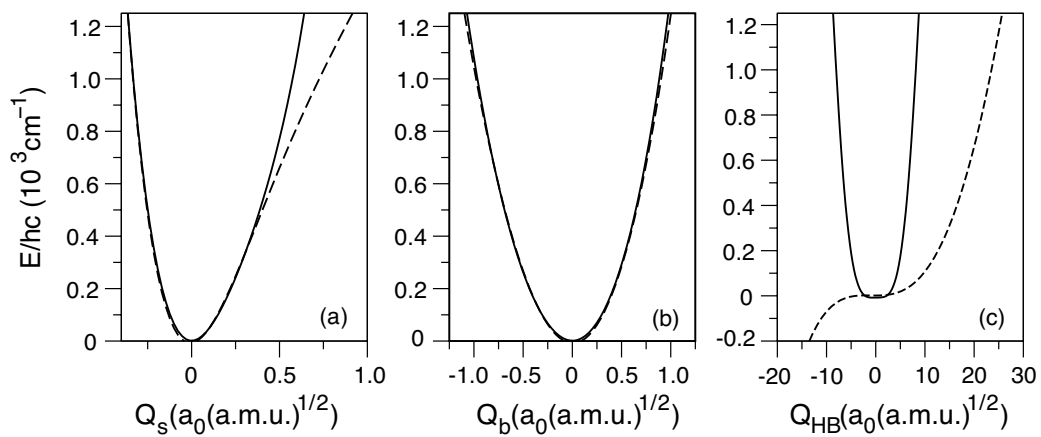}
\caption{Potential energy curves along modes $\ms$ (a), $\mb$ (b), and $\nu_{\hb}$
(c). The solid lines correspond to the potentials obtained on a grid, whereas the dashed lines
stand for the Taylor expansions up to the 4$^{\rm th}$ order according to Eq. \re{taylor}.}
\label{fig:afc}
\end{center}
\end{figure}

The question arises whether the fourth-order approximation, Eq. \re{taylor}, is justified for describing the selected modes.
In order to scrutinize this point we plot in Fig. \ref{fig:afc} exemplarily the potentials along $Q_{\rm s}$, $Q_{\rm b}$, and $Q_{\hb}$ as obtained from the fourth-order expansion (dashed line) and from a numerical evaluation on a  grid (solid line). According to panel (a), the anharmonic force field description of $\ms$ reproduces the exact potential for small displacements, but the two curves start to deviate above $\sim$4000 \cm. The anharmonicity of the bending mode, $\mb$, is not very pronounced what is reflected in a good agreement between the two curves up to 12000 \cm in panel (b). The fourth order description breaks down completely for the low-frequency mode,  $\nu_{\hb}$,  as seen in panel (c). The anharmonicity of  $\go$ and $\gt$ is only modest and  fourth-order expansions and grid based potentials essentially agree over the considered range. Nevertheless, we have chosen to use the obtained grid-based data for representing all one mode potentials $V_i^{(1)}(Q_i)$. Further, the following two-mode potentials have been generated on a numerical grid: $V^{(2)}_{\rm s,b}(Q_{\rm s},Q_{\rm b})$, $V^{(2)}_{\rm s,{\hb}}(Q_{\rm s},Q_{\hb})$, and $V^{(2)}_{b,{\hb}}(Q_{\rm b},Q_{\hb})$. This selection is motivated by the focus which is on the OH-stretching region of the spectrum where (i) the Fermi-resonance type coupling between the stretching fundamental and the bending overtone is of particular importance (see Tab. \ref{tab:afc}) and (ii) low-frequency wave packet motion can be triggered by OH stretching excitation. All other correlations up to the four-mode term have been included in terms of the calculated force constants.

The resulting Hamiltonian has been diagonalized assuming a diagonal kinetic energy operator in a step-wise procedure as follows:
First, the uncoupled one-dimensional Hamiltonians have been diagonalized (note that the coordinates are mass-weighted)
\be
\left[-\frac{\hbar^2}{2}\frac{\partial^2}{\partial Q_i^2}+ V_i^{(1)}(Q_i)\right] \ket{\phi_n^{i}} = E_n^{i} \ket{\phi_n^{\nu_i}}
\ee
by using the Fourier-Grid-Hamiltonian (FGH)  method \cite{marston89:3571} (Grid parameters: $Q_{\rm s}$ - 128 points in [-0.9:1.7], $Q_{\rm b}$ - 64  points in  [-2.0:2.0], $Q_{\go}$ and $Q_{\gt}$ - 64 points in [-2.1  2.1],   $Q_{\hb}$ - 128 points in [-8.0:8.0] (intervals in $a_{0}\sqrt{\rm a.m.u.}$)). Second, from these zero-order functions $\{\ket{\phi_n^{i}} \}$, two product bases are formed for the modes  ($\ms,\mb$) and  ($\go ,\gt$) including 5$\times$ 7 and 7$\times$ 7 functions, respectively. The respective 2-mode Hamiltonians are diagonalized and the lowest 5 and 20 eigenfunctions are used to span a product basis for the diagonalization of the 4-mode Hamiltonian at the equilibrium geometry of the low-frequency mode $\nu_{\hb}$. Notice that the parameters for the successive diagonalization have been chosen such as to obtain converged eigenstates for transitions up to about 3600 \cm.

Due to the time scale separation between the four fast modes $\ms$, $\mb$, $\go$, and $\gt$ and the mode $\nu_{\hb}$, the resulting eigenvalues, which span a diabatic basis, can be used for the characterization of the nature of the fast modes' states. Denoting the diabatic basis as $\{\ket{\alpha^{\rm diab}} \}$, these states obey the following Schr\"odinger equation:
\be
\Big[ \sum\limits_{i={\rm s,b},\gamma_1,\gamma_2}  \Big[-\frac{\hbar^2}{2}\frac{\partial^2}{\partial Q_i^2}+ V_i^{(1)}(Q_i)\Big]   + V^{\rm (c)}({\bf Q}_{\rm fast}, Q_{\hb}=0)
\Big]\ket{\alpha^{\rm diab}} = E^{\rm diab}_{\alpha} \ket{\alpha^{\rm diab}}.
\ee
Here, ${\bf Q}_{\rm fast}$ comprises the four fast modes,  whereas $V^{\rm (c)}({\bf Q}_{\rm fast}, Q_{\hb}=0)$ stands for the 
couplings between them. Expressed in the  basis of the uncoupled single mode states the four-dimensional diabatic states read
\be
\ket{\alpha^{\rm diab}} = \sum\limits_{ijkl} C_{ijkl}^{\alpha}
\ket{\phi_i^{\rm s}} \ket{\phi_j^{\rm b}}
\ket{\phi_k^{\go}} \ket{\phi_l^{\gt}}
\ee
with the expansion coefficients $C_{ijkl}^{\alpha}$. In the simulations below we have included the lowest 25 diabatic states; for an assignment see Tab. \ref{tab:eigenstates}. %

Having defined the diabatic states, the total Hamiltonian in the diabatic representation obtains the following form
\bea 
H_{\rm diab} &=& \sum\limits_{\alpha\beta} \Big[
\delta_{\alpha\beta} \Big( -\frac{\hbar^2}{2}\frac{\partial^2}{\partial Q_{\hb}^2} + E^{\rm diab}_{\alpha} +
V_{\alpha\alpha}(Q_{\hb}) \Big) \nonumber \\
&+& ( 1 -
\delta_{\alpha\beta} ) V_{\alpha\beta}(Q_{\hb}) \Big]
\ket{\alpha^{\rm diab}} \bra{\beta^{\rm diab}}
\label{eq-pmme-totalH}
\eea
where the matrix elements between the diabatic states are given by
\be 
\label{eq:vab}
V_{\alpha\beta}(Q_{\hb}) = \bra{\alpha^{\rm diab}} V_{\hb}^{(1)}(Q_{\hb}) +  V^{\rm (c)} ({\bf Q}_{\rm fast}, Q_{\hb}) - V^{\rm (c)} ({\bf Q}_{\rm fast}, Q_{\hb}=0) \ket{\beta^{\rm diab}} \, . 
\ee

In a final step the Hamiltonian, Eq. \re{eq-pmme-totalH}, is diagonalized after introducing zero order states of the low-frequency mode with respect to the different diabatic states' potential energy curves. Labeling the resulting 5D eigenstates as $\{ \ket{\alpha} \}$ the IR stick spectrum can be expressed as \cite{may04}
\be
I^{(\rm stick)}(\omega) \propto \omega \sum_{i=x,y,z}\sum\limits_{\alpha\beta} P_{\alpha} |d^{(i)}_{\alpha\beta}|^2 \delta (\omega-\omega_{\alpha\beta}) \, .
\label{eq:stick}
\ee
Here, $P_{\alpha}$ is the thermal distribution function, and $d^{(i)}_{\alpha\beta}$ and
$\omega_{\alpha\beta}$ stand for the $i$th  transition dipole moment component and the transition frequency, respectively. For the dipole moment vector $\mathbf{ d}$ we
have incorporated its three components, $d^{(i)}$, by expressing them on one-dimensional grids along $Q_{\rm s}$ and $Q_{\rm b}$. In other words, only selected one-mode contributions to the dipole moment function are considered.

\subsection{QM/MM Approach}
%
PMME solvated in CCl$_4$ is treated on the basis of the hybrid QM/MM method provided by the CPMD/Gromos
interface~\cite{laio02:6941}.  The QM/MM separation has been done in the solute/solvent fashion.  The solvent molecules are coped 
with Gromos adopting the Gromos96 force field~\cite{scott99:3596}. The PMME molecule is dealt with CPMD \cite{cpmd} using
the Becke exchange and Lee-Yang-Par correlation  functional (BLYP)~\cite{becke88:3098,lee88:785}.
Further, the Troullier-Martins (TM) pseudopotential~\cite{troullier91:1993} with  a wavefunction cutoff of 70 Ry is adopted to describe the interaction between  the core and valence electrons in the quantum region.  For the fictitious electron mass we use 400 a.u.

In the simulation, one PMME is solvated in 769 $\textrm{CCl}_4$ molecules
and the solution is put into a box with dimensions
50\AA$\times$50\AA$\times$50\AA.
Before the QM/MM simulation, a classical solvent equilibration with Gromos 
is carried out for 1 ns at 300 K with the solute fixed using the SHAKE algorithm.
The quantum part is placed in a $15.88~\textrm{\AA}\times12.70~\textrm{\AA}\times19.05~\textrm{\AA}$ box. 
The trajectory run is performed at 300 K with a Nos\'e-Hoover chain thermostat 
and by using a time step of 2 a.u. (0.048 fs). The total simulation time is 16 ps, which 
takes about 40 days on a cluster with 24 2.66 GHz CPUs. The first 0.5 ps has been assigned for equilibration of the QM/MM box.

The resulting trajectory has been analyzed putting emphasis on the geometry of the HB. Further, dipole and velocity  autocorrelation functions were calculated to give the IR absorption spectrum and the density of states, respectively.
The IR absorption coefficient of the solute, $I^{\rm d-d}(\omega)$, can be calculated from the dipole-dipole correlation function \cite{may04}
\begin{equation}
\label{eq:IR}
I^{\rm d-d}(\omega)\propto \omega {\rm Re} \int^\infty_{0} dt e^{i\omega t} \langle \mathbf{ d}(t)\cdot\mathbf{ d}(0)\rangle \,,
\end{equation}
whereas the density of states is given by~\cite{dickey69:1407} 
\begin{equation}
\label{eq:DOS}
N(\omega)=\int dt e^{i\omega t} \sum_i \langle \mathbf{v}_i(t)\cdot\mathbf{v}_i(0)\rangle \, .
\end{equation}
Further, we have calculated anharmonic OH-stretching and bending  frequencies along the trajectory by generating two-dimensional snap-shot potential energy surfaces for an otherwise fixed geometry of solute and solvent \cite{jezierska07:205101,stare08:1576}. Anharmonic vibrational eigenstates have been obtained using  sine basis sets \cite{locker71:1756} and including the kinetic energy coupling between the coordinates. The potential  is obtained by moving the H atom on a two-dimensional plane defined by the atoms C8-O1-H. In the calculation 24 snap-shots are sampled starting at the production part of the trajectory with an interval of 2000 steps. For each snap-shot, the O1-H bond stretching is sampled with 7 grid points in the region 0.7-2.0~\AA. The  O1-H   bending within the C8-O1-H plane is sampled with 7 points from -45$^\circ$ to 45$^\circ$. 

%
\section{Results and Discussion}
\label{sec:res}
%
\subsection{Analysis of Gas Phase IR Spectrum}
\label{sec:res:gas}
The decomposition of the diabatic states of the fast modes is given in Tab. \ref{tab:eigenstates} for the range 2200-3500 \cm . 
Most of the states correspond to almost pure combination and overtone transitions. Exceptions are the states  $\ket{\alpha^{\rm diab}=16}$ and $\ket{\alpha^{\rm diab}=20}$. Both are mixed states with respect to the bending overtone and the stretching fundamental transitions. They are found at 2853 and 3044 \cm, respectively. Interestingly, both states have contributions from combinations of the $\delta_{\rm OH}$ fundamental and the $\gamma_{1/2}$ overtone transitions. 

\begin{figure}
\begin{center}
\includegraphics[width=0.8\textwidth]{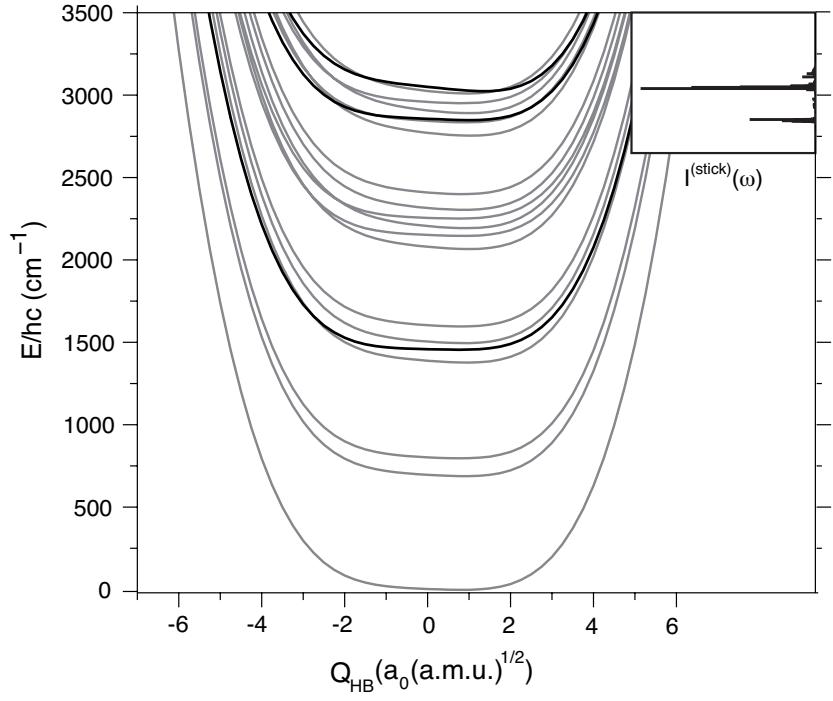}
\caption{Diabatic potential energy curves of the lowest 20 states. The thick solid lines mark the states which are dominated by the bending fundamental, $\ket{\alpha^{\rm diab}=5}$, its first overtone $\ket{\alpha^{\rm diab}=16}$ as well as the stretching fundamental, $\ket{\alpha^{\rm diab}=20}$, transition. The insert shows the stick spectrum according to Eq. \re{eq:stick}.}
\label{fig:diab}
\end{center}
\end{figure}

The diabatic potential energy curves $V_{\alpha\alpha}(Q_{\hb})$, Eq. \re{eq:vab},  are depicted in Fig. \ref{fig:diab}. The potential curves for the diabatic states  $\ket{\alpha^{\rm diab}=16}$ and $\ket{\alpha^{\rm diab}=20}$ (marked in black) are found to be embedded in the manifold of curves originating from nearby states. Since there are no symmetry restrictions, the low-frequency mode in principle will couple all states in this region of the present model. This gives rise to the stick spectrum, Eq. \re{eq:stick}, shown in the insert of Fig. \ref{fig:diab}. It is dominated by the above mentioned Fermi-resonance, but also contains the signature of the progression with respect to the low-frequency mode on the blue side of the $\ms$ dominated peak. 
Judging the intensities of te various peaks, it should be kept in mind that the dipole surface is described by selected one mode terms only.
Note that this progression is the origin of the wave packet motion of the low-frequency HB mode which has been observed in Ref. \cite{madsen02_909}. Needless to point to the pronounced anharmonicity of $\nu_{\hb}$ which together with the many diabatic levels coupled in this spectral range makes the dynamical problem considerably more complicated than that of shifted harmonic oscillators.
\subsection{Hydrogen Bond Geometry in Solution}
\begin{figure}[ht]
\centering
\includegraphics[width=14cm]{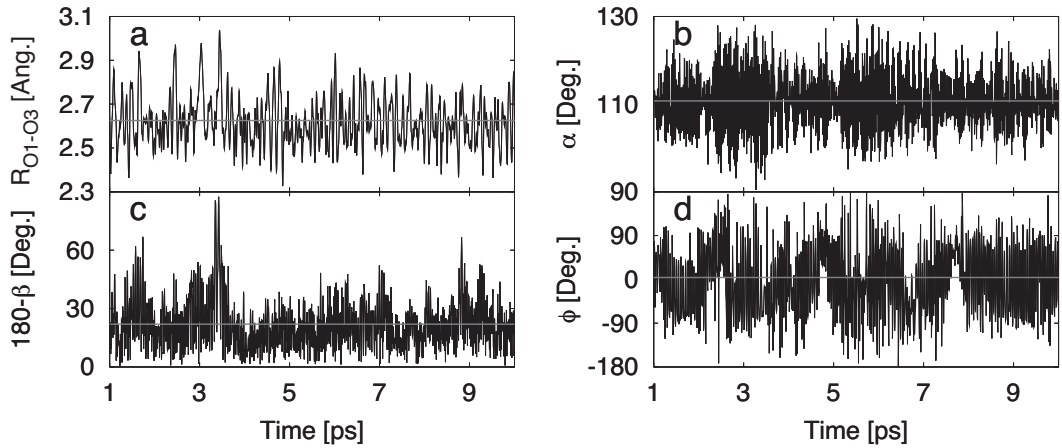}
\caption{Geometric parameters of the HB in PMME along a representative part of the trajectory :
(a) HB length, $R_{\rm O1-O3}$,  (b) bending angle $\alpha= \angle({\rm C8,O1,H})$,
(c) out-of-line motion of the hydrogen measured by the difference
between 180$^\circ$ and the HB angle $\beta=\angle ({\rm O1,H,O3})$,
(d) dihedral angle of the HB $\phi=\angle ({\rm C8,O1,H,O3})$.
The horizontal line in each panel indicates the corresponding time average over the whole production trajectory.
}
\label{fig:geo}
\end{figure}

The geometry of the HB along the trajectory will be characterized by the HB length, $R_{\rm O1-O3}$, the angle, $\beta=\angle ({\rm O1,H,O3})$,  the bending angle of the OH vibration, $\alpha= \angle({\rm C8,O1,H})$, and the dihedral angle $\phi=\angle ({\rm C8,O1,H,O3})$; for labeling see Fig. \ref{fig:struct}. The change of these parameters along part of  the production trajectory is given in Fig. \ref{fig:geo}. Note that the OH bond length itself does not change appreciably, its average value is 1.02~\AA{} and the variation is 0.04~\AA. In other words, this system does not support any Hydrogen transfer similar to the prediction of the gas phase calculations \cite{paramonov01:205}. From Fig. \ref{fig:geo}a we obtain the average HB length of $R_{\rm O1-O3}= 2.63$\AA{} and a RMS variation of 0.11 \AA. As compared with the gas phase DFT/B3LYP 6-31+G(d,p) geometry the HB is on average elongated by 0.07\AA{} \cite{paramonov01:205}. Large elongations correlate with strong deviations of the HB from linearity as seen in panel (c) of Fig. \ref{fig:geo}, e.g. for $\sim$1.5 ps and $\sim$3.2 ps. The average deviation from linearity is 24$^\circ$. The time evolution of the OH bending angle $\alpha$ is given in Fig. \ref{fig:geo}b, its average and RMS variation are 111$^\circ$ and 6$^\circ$, respectively. Finally, we focus on the planarity of the HB with respect to the dihedral angle $\phi$. Note that the gas phase structure of PMME has been calculated to be overall nonplanar \cite{paramonov01:205}.  The gas phase DFT value is obtained as $\phi=-51^\circ$. In condensed phase the average and RMS variation are 15$^\circ$ and 65$^\circ$, respectively.  The distribution of $\phi$ values is, however, bimodal with the main peak at 57$^\circ$ and a 40\% smaller peak at -53$^\circ$. Thus, in comparison with the gas phase PMME assumes more often a structure where carboxyl and ester groups are twisted with respect to each other. 
This fact also indicates the limitations of the gas phase model presented above.
Finally, we point out that the frequently observed  changes from 180 to -180$^\circ$ and vice versa are indicative of  linear HB geometries.
\subsection{IR Spectrum in Solution}
\label{sec:res:cond}
\begin{figure}[ht!]
\centering
\includegraphics{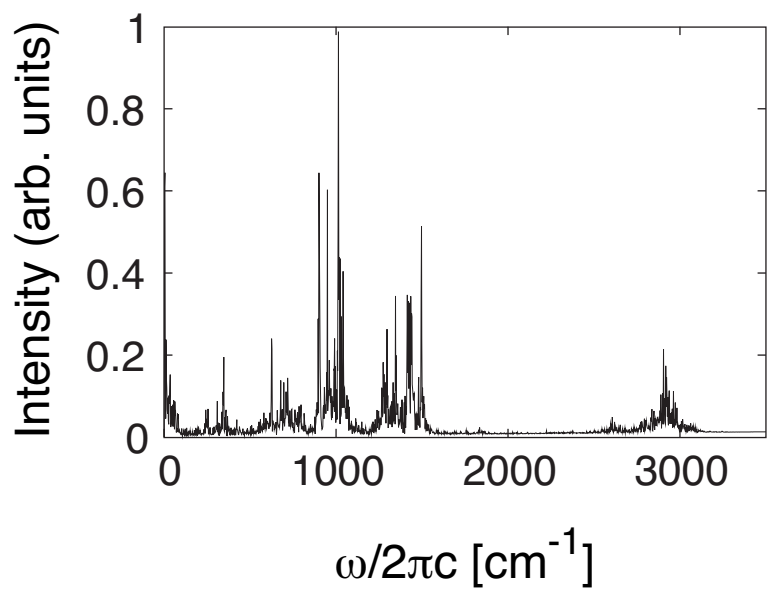} \\
\caption{
Total IR spectrum of PMME in CCl$_4$ solution at 300K as obtained from Eq. \re{eq:IR} using the 15.5 ps production QM/MM trajectory. Note that the contribution of solvent absorption is not included since the dipole autocorrelation function is calculated for the QM part only.
}
\label{fig:IR}
\end{figure}

\begin{figure}[ht!]
\centering
\includegraphics[width=0.8\textwidth]{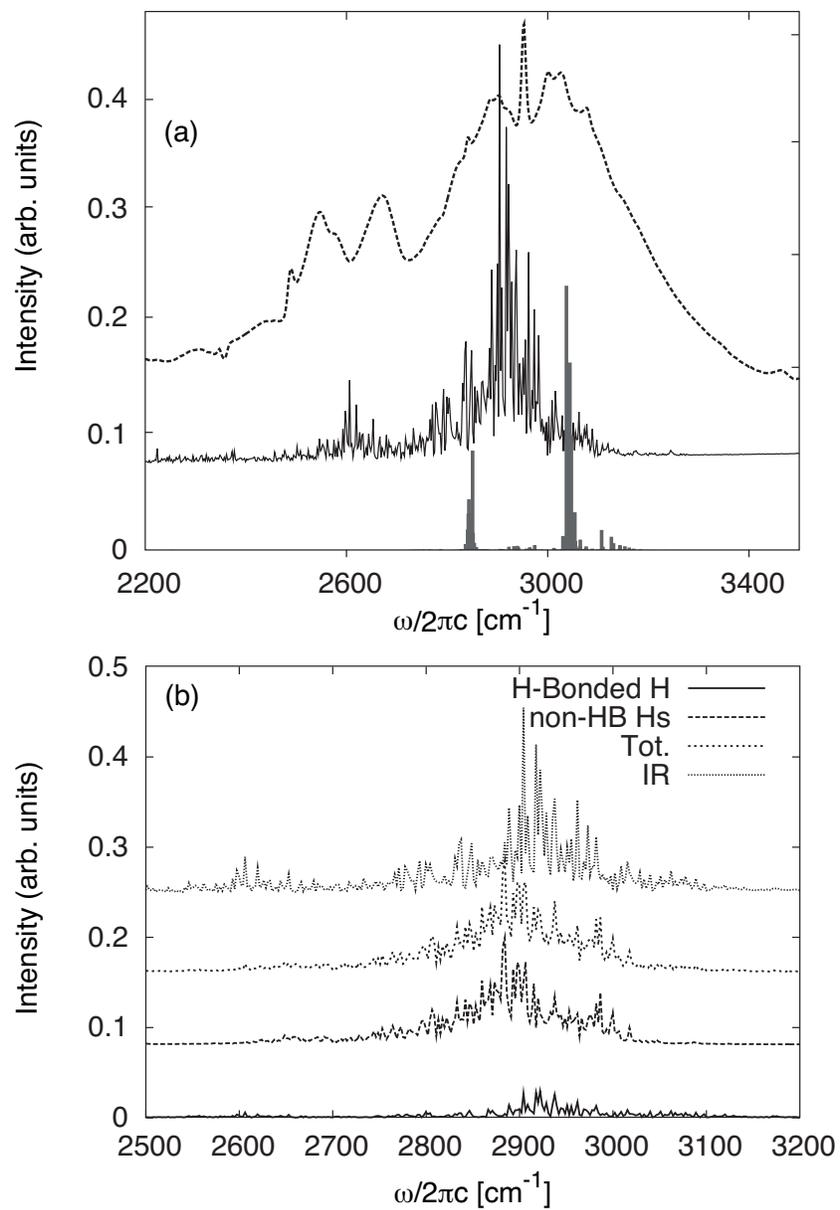} 
\caption{(a) Experimental (dashed) \cite{madsen02_909}, QM/MM (solid), and gas phase (sticks) IR spectra. 
(b) Comparison between the density of states, Eq. \re{eq:DOS}, and the IR spectrum as indicated in the figure key.
}
\label{fig:specana}
\end{figure}

The total IR spectrum according to Eq. \re{eq:IR} is shown in Fig. \ref{fig:IR}. Our focus will be on the OH-stretch region around 3000 \cm which is shown enlarged in Fig. \ref{fig:specana}.  Panel (a) compares the calculated IR spectrum with the experimental result of Ref. \cite{madsen02_909}. The experimental spectrum exhibits a broad band covering the range from   $\sim$2750 to $\sim$3250 \cm. Comparison with the spectrum of the deuterated species \cite{nibbering07:619} shows that the  sharp feature around 2950 \cm must be due to some CH stretching vibration which is not coupled to the HB motion. The two features to the left and right of this peak survive deuteration as well. The two peaks around $\sim$2600 \cm, on the other hand, disappear upon deuteration and should be related to the HB. 

The calculated IR spectrum, although not reproducing the full width of the measured band, compares rather well with experiment. This concerns the position of the main band, but also the appearance of some intensity around 2600 \cm. The assignment of the spectrum can be guided by comparison with the stick spectrum of the gas phase model which is also shown in Fig. \ref{fig:specana}a. Essentially, the two Fermi-resonance lines are in the relevant spectral region of the main experimental band, giving evidence that the latter is indeed containing these transitions. Its width is likely to originate from the fluctuations of the HB parameters as given in Fig. \ref{fig:geo}. In addition the position of the stick spectrum agrees nicely with that coming from the condensed phase simulation. As far as the features around 2600 \cm are concerned we can only conclude that the respective transitions are not due to excitations of those modes which are part of the 5D model, see Fig. \ref{fig:modes}.

Another means for gaining insight into the nature of the motions contributing to the IR spectrum is an analysis in terms of the density of states due to certain groups of atoms. Respective results are given in Fig. \ref{fig:specana}b. Overall the width of IR spectrum in the considered range  is comparable to the density of states. The latter is, however, dominated by contributions of non-H-bonded Hydrogen motions. Contributions from the Hydrogen of the HB are in the range between $\sim$ 2850 and 3000 \cm which is in accord with the discussion of panel (a). Interestingly there is some intensity of $N(\omega)$ around 2600 \cm, i.e. the mentioned features in the IR spectrum are indeed likely to be due to the dynamics of the HB as suggested by deuteration studies.

\begin{figure}[ht!]
\centering
\includegraphics{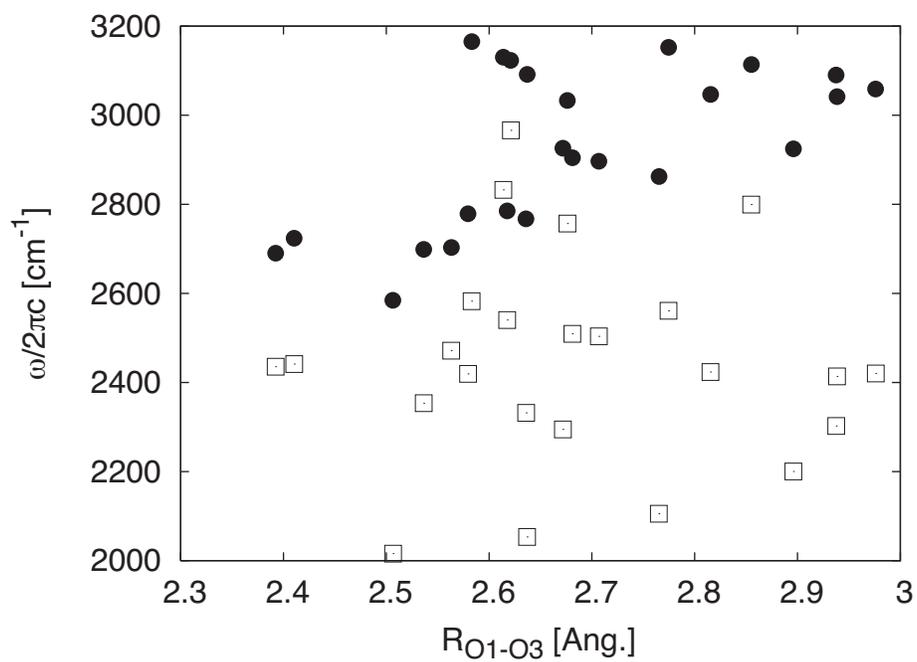} 
\caption{The correlation between the HB bond length, $R_{\rm O1-O3}$, and 
the OH-stretching fundamental transition frequency (bullets)
as well as the first   bending overtone (hollow squares). For definition of coordinate see text.
}
\label{fig:corr}
\end{figure}

The quantum mechanical calculation of the linewidth in condensed phase requires to have at hand the fluctuations of the transition frequencies of the high frequency modes due to their interaction with the surrounding modes \cite{mukamel95}. In Ref. \cite{yan08:230} we have used an empirical mapping between HB distances and transition frequencies to predict the linewidth of the NH-stretching transition in a solvated adenine-uracil base pair. This mapping was based on available crystal structure data \cite{novak74:177}, see also Ref. \cite{bratos04:197}. However, no such empirical correlation is available for PMME type of molecules. An alternative could be the explicit calculation of potential energy curves for selected coordinates on-the-fly along the trajectory \cite{jezierska07:205101,stare08:1576}. Figure \ref{fig:corr} shows results of respective calculations for the  Fermi-resonance states, i.e. the dependence on the bending overtone and the stretching fundamental transition on the HB length, $R_{\rm O1-O3}$. For the $\ms$ mode one can recognize the expected correlation, i.e. the increase of the transition frequency with  $R_{\rm O1-O3}$. Note that this holds irrespective of the nonlinearity and wide variability of the HB parameters seen in Fig. \ref{fig:geo}. For the bending overtone, on the other hand,  no such correlation is discernible. The reason can be found in the difficulty to define the coordinate for this bending motion due to the large flexibility of the structure which mixes different types of bending motions, a fact that cannot be captured by the present definition of coordinates within the two-dimensional model.
%
\section{Summary}
\label{sec:sum}
%
The IR absorption spectrum of  PMME in the spectral range of the Hydro\-gen-bonded OH-stretching vibration has been investigated using a five-dimen\-sional gas phase quantum mechanical model as well as condensed phase classical molecular dynamics simulations on the basis of the QM/MM approach. The gas phase model predicts specific anharmonic coupling patterns between the high-frequency OH stretching vibration and lower-frequency bending modes. In particular it was found that the region around 3000 \cm contains a Fermi resonance between the $\ms$-stretching fundamental and the first $\mb$-bending overtone vibration with considerable oscillator strength redistribution. The effect of low-frequency HB motion has been studied within the picture of potential energy curves defined for the diabatic states of the higher-frequency modes. These curves are markedly anharmonic and form a dense manifold of coupled states in the 3000 \cm range. In the IR spectrum the low-frequency mode gives rise to a vibrational progression along with the $\ms$ fundamental transition.  
These finding are similar to the case of deuterated PMME \cite{kuhn02:7671}. 

The condensed phase simulations of the IR spectrum  gave a broad band in the 3000 \cm region whose position is not only in agreement with the main experimental band. In addition there is a weak band around 2600 \cm coinciding with an experimental feature. Analysis in terms of the density of states led to the conclusion that this band has contributions from HB motions. 
Comparing gas and condensed phase IR spectra we found an overall good agreement as far as the spectral range is concerned. However, the trajectory simulations also indicate that in solution a different configuration may also be present and contribute to the spectrum.
Finally, we have explored the possibility to correlate $\ms$ fundamental and $\mb$ overtone excitation to the length of the HB. While there appears to be a reasonable correlation with respect to $\ms$, the scatter in the $\mb$ points is considerable, thus pointing to the difficulty of defining the bending motion in terms of a single coordinate.
\section*{Acknowledgments}
We gratefully acknowledge financial support by the Deutsche Forschungsgemeinschaft (Sfb450 (M.P.,G.K.) and project Ku952/5-1 (Y.Y.,O.K.)).
%


\begin{table} [h!]
\begin{center}
{
\doublerulesep 0pt
\begin{tabular}{lrp{1.0cm}|l|r|}
\cline{1-2}\cline{4-5}
\cline{1-2}\cline{4-5}
\multicolumn{1}{|l|}{$i,j,k$}   & \multicolumn{1}{|r|}{$K_{ijk}$ in \cm}
                                        & &
                               $i,j,k,l$       & $K_{ijkl}$ in \cm \\
\cline{1-2}\cline{4-5}
\multicolumn{1}{|l|}{$   s,   s,   s$}   & \multicolumn{1}{|r|}{ -2867}
                                        & &
                               $   s,   s,   s,   s$     &  1853         \\
\multicolumn{1}{|l|}{$   b,   b,   b$}   & \multicolumn{1}{|r|}{   -130}
                                        & &
                               $   b,   b,   b,   b$     &   176          \\
\multicolumn{1}{|l|}{$ \hb, \hb, \hb$}   & \multicolumn{1}{|r|}{  -20}
                                        & &
                               $ \hb, \hb, \hb, \hb$     &   384          \\
\cline{1-2}
\multicolumn{1}{|l|}{$   s,   s,   b$}   & \multicolumn{1}{|r|}{  62}
                                        & &
                               $ \go, \go, \go, \go$     &    40        \\
\multicolumn{1}{|l|}{$   s,   b,   b$}   & \multicolumn{1}{|r|}{  498}
                                        & &
                               $ \gt, \gt, \gt, \gt$     &    41          \\
\cline{1-2}
\cline{4-5}
\multicolumn{1}{|l|}{$   s,   s, \hb$}   & \multicolumn{1}{|r|}{  -73}
                                        & &
                               $   s,   s,   s,   b$     &    -83          \\
\multicolumn{1}{|l|}{$   b,   b, \hb$}   & \multicolumn{1}{|r|}{   15}
                                        & &
                               $   s,   s,   b,   b$     &  -555          \\
\cline{1-2}
\multicolumn{1}{|l|}{$   s,   s, \go$}   & \multicolumn{1}{|r|}{   46}
                                        & &
                               $   s,   b,   b,   b$     &   73          \\
\multicolumn{1}{|l|}{$   s,   s, \gt$}   & \multicolumn{1}{|r|}{   50}
                                        & &
                               $   s,   s, \hb, \hb$     &  -275          \\
\cline{1-2}
\multicolumn{1}{|l|}{$   b,   b, \go$}   & \multicolumn{1}{|r|}{   -7}
                                        & &
                               $   s,   s, \hb, \go$     &    74          \\
\multicolumn{1}{|l|}{$   b,   b, \gt$}   & \multicolumn{1}{|r|}{   -2}
                                        & &
                               $   s,   s, \hb, \gt$     &   100          \\
\cline{1-2}
\multicolumn{1}{|l|}{$   b, \go, \go$}   & \multicolumn{1}{|r|}{    -8}
                                        & &
                               $   s,   s, \go, \go$     &   -35          \\
\multicolumn{1}{|l|}{$   b, \gt, \gt$}   & \multicolumn{1}{|r|}{    -9}
                                        & &
                               $   s,   s, \gt, \gt$     &   -80          \\
\multicolumn{1}{|l|}{$   b, \go, \gt$}   & \multicolumn{1}{|r|}{   -11}
                                        & &
                               $   s,   s, \go, \gt$     &   -47          \\
\cline{1-2}\cline{4-5}
                      &                 & &
                               $   b,   b,   b, \hb$     &    -14          \\
                      &                 & &
                               $   b,   b, \hb, \hb$     &    -30          \\
                      &                 & &
                               $   b, \hb, \hb, \hb$     &    -22          \\
\cline{4-5}
\end{tabular}
}
\caption{Selected cubic (left) and quartic (right) anharmonic coupling constants for the chosen 5D gas phase model.}
\label{tab:afc}
\end{center}
\end{table}

\clearpage


\begin{table} 
\begin{center}
{ \doublerulesep 0pt
\begin{tabular}{||r|r|r|r||}
\hline
\hline
 level & $E/hc$ (\cm) & ($v_{\rm s}$,$v_{\rm b}$,$v_{\go}$,$v_{\gt}$) & $C_{ijkl}^{\alpha}$   \\
\hline
   10  &         2205 &  (0,0,1,2) &  -0.978 \\
         &                 &  (0,1,0,1) & 0.109 \\
         \hline
   11  &         2253 &  (0,1,1,0) &  0.987\\
   \hline
   12  &         2312 &  (0,0,2,1) &  0.969 \\
        &                  &  (0,0,3,0) & -0.139 \\
        &                  &  (0,0,1,0) &  0.105 \\
    \hline
   13  &         2407 &  (0,0,3,0) &  -0.982 \\
        &                  &  (0,0,2,1) & -0.135 \\
    \hline
   14  &         2779 &  (0,0,0,4) & -0.988 \\
   \hline
   15  &         2844 &  (0,1,0,2) &  0.914 \\
        &                  &  (0,2,0,0) & -0.318\\
        &                  &  (1,0,0,0) &  0.175\\
\hline
   {\bf 16}  &         {\bf 2853} &  (0,2,0,0) & 0.803 \\
       &              &  (1,0,0,0) &  -0.449 \\
       &              &  (0,1,0,2) &   0.362\\
\hline
   17  &         2910 &  (0,0,1,3) &  0.970 \\
        &                  &  (0,1,0,2) &  -0.115 \\
 \hline       
   18  &         2956 &  (0,1,1,1) & -0.975 \\
        &                  &  (0,0,2,2) & -0.110\\
  \hline
   19  &         3025 &  (0,0,2,2) &  0.954  \\
        &                 &   (0,0,3,1) &  0.167 \\
        &                 &   (0,1,1,1) &  -0.129\\
        &                 &   (0,0,2,3) & -0.110\\
\hline
   {\bf 20}  &         {\bf 3044} &  (1,0,0,0) & -0.853 \\
       &              &  (0,2,0,0) & -0.472 \\
       &              &  (0,1,2,0) & -0.189 \\
\hline
   21  &         3055 &  (0,1,2,0) & -0.962 \\
         &                 &  (1,0,0,0) & 0.162 \\
         &                 &  (0,2,0,0) & 0.107 \\
\hline   
   22  &         3126 &  (0,0,3,1) &  -0.940 \\
   23  &         3217 &  (0,0,4,0) & -0.963 \\
   24  &         3477 &  (0,0,0,5) &  0.984 \\
\hline\hline
\end{tabular}}
\caption{Energies of the diabatic states in the interval from 2200 to 3500 \cm assignment, expressed in terms of the uncoupled anharmonic modes. The last column contains the expansion coefficients ($>$ 0.1 for states 10-21 and only leading coefficients for states 22-24).}
\label{tab:eigenstates}
\end{center}
\end{table}
%
\clearpage\newpage

\clearpage
\newpage

\clearpage
\newpage
\clearpage
\newpage

\clearpage\newpage
\clearpage\newpage
\clearpage\newpage
\clearpage\newpage

\end{document}